\documentclass[12pt]{iopart}

\usepackage{graphicx}
\usepackage{dcolumn}
\usepackage{cite}

\begin{document}

\title{Searching for structure beyond parity in the two-qubit Dicke model}

\author{B. M. Rodr\'{\i}guez-Lara}
\address{Instituto Nacional de Astrof\'{i}sica, \'{O}ptica y Electr\'{o}nica \\ Calle Luis Enrique Erro No. 1, Sta. Ma. Tonantzintla, Pue. CP 72840, M\'{e}xico}
\ead{bmlara@inaoep.mx}

\author{S. A. Chilingaryan}
\address{Departamento de F\'{i}sica, Universidade Federal de Minas Gerais, Caixa Postal 702, 30123-970, Belo Horizonte, MG, Brazil}

\author{H. M. Moya-Cessa}
\address{Instituto Nacional de Astrof\'{i}sica, \'{O}ptica y Electr\'{o}nica \\ Calle Luis Enrique Erro No. 1, Sta. Ma. Tonantzintla, Pue. CP 72840, M\'{e}xico}

\begin{abstract}
We try to classify the spectrum of the two-qubit Dicke model by calculating two quantum information measures of its eigenstates: the Wooters concurrence and the mutual quantum information.
We are able to detect four spectral sets in each parity subspace of the model: 
one set is regular and given by the product of a Fock state of the field times the singlet Bell state of the qubits; the rest are fairly regular and related to the triplet states of the Bell basis.
The singlet states become trapping states when we couple the Dicke model to an environment of harmonic oscillators, making them candidates for generating maximally entangled states in experimental realizations of  ion trap quantum electrodynamics (QED) and circuit QED.
Furthermore, they are robust and survive the inclusion of driving and dipole-dipole interactions, pointing to their use for storing quantum correlations, and it is straightforward to provide a generalization of these trapping states to the Dicke model with even number of qubits.
\end{abstract}


\maketitle

\section{Introduction} \label{sec:S1}

The presence of regularity in energy or quasi-energy spectra points to the existence of symmetries that translate into conserved dynamical variables \cite{Peres1984p1711}.  
In any generic system the spectrum consists of a regular and an irregular part \cite{Percival1973p229,Carlo1998p5397}.
If the classical analogue of a quantum model is fully integrable, then a set of quantum numbers ascribed to the energy levels ought to exist.
For fully non-integrable classical systems, their quantum analogue spectra should be irregular and the energy levels cannot be labeled with quantum numbers related to conserved dynamical variables.
Sometimes the system may have a partial dynamical symmetry \cite{Leviathan2011p93}, and then one of three cases may arise: (i) some of the proper states conserve a certain constant of motion, (ii) all of the proper states conserve part of a constant of motion, or (iii) some of the proper states conserve a part of a constant of motion.

We are interested in the symmetries of two identical qubits interacting with a quantized field under the Dicke model \cite{Dicke1954p99} with the inclusion of the so-called counter-rotating terms. 
This two-qubit full Dicke model has a well known parity symmetry \cite{Chilingaryan2013p335301,Peng2013} but no other fully conserved constant of motion has been found for it.
It can be realized experimentally with solid state devices \cite{Liu2006p052321,Devoret2007p767} and trapped ions \cite{Liu2007p1967,Genway2013}. 
Our interest in the symmetries is twofold.  
First,  finding the conserved dynamical variables of a quantum model helps in constructing trapping states for them.
Trapping states are the stationary states of a system that do not evolve in the presence of losses.
In the single-qubit case in the weak coupling regime, trapping states can be used to generate macroscopic superposition states of the field    \cite{Slosser1990p233,Weidinger1999p3795}. 
Second, once all the conserved dynamical variables are found, it is straightforward to provide an analytic time evolution operator which can help in the design of two-qubit quantum gates outside the weak coupling regime \cite{Wang2012p014031}.
To the best of our knowledge, while the spectra and integrability of a single-qubit \cite{Braak2011p100401} and a three-qubit \cite{Braak2013p224007} Dicke model have been studied, a detailed statistical description of the spectra for the two-qubit model does not exist and nothing has been said about its trapping states. 
Some of us have explored analytic methods for diagonalizing the two-qubit Rabi model \cite{Chilingaryan2013p335301} and another analytic description of its spectra has been discussed during the reviewing of this work \cite{Peng2013}, the ground state of the two-qubit Dicke model in the ultra-strong coupling regime has been recently approximated by using a variational method \cite{Lee2013p015802}, and the dynamics of quantum correlations of two qubits coupled to a bath with a Lorentzian spectrum and independent reservoirs has been given numerically \cite{Wang2013p103020}.

Here, we study the structure of the spectrum and eigenstates of the two-qubit Dicke model by means of Peres suggestion for finding conserved quantities in model Hamiltonians \cite{Peres1984p1711}.
We use measures of entanglement from quantum information theory: the Wootters concurrence and the quantum mutual information of the qubit ensemble, which allow us to propose a classification of the two-qubit Dicke model spectrum into four fairly regular spectral classes.
One of these classes is related to the singlet state in the Bell basis which can act as a trapping state for the qubit part if the system is open due to its well known properties \cite{Bell1964p195}.  
In order to provide an extra example for the method for classifying the spectra, we also discuss a variation of the Dicke model realizable in ion traps that yields trapping states of the form of other Bell states.
Then, we generalize these classes of trapping states to systems composed of even numbers of qubits.
Bell states are maximally entangled and their generation and preservation is a crucial sought-after goal for quantum communication and information purposes; cf. \cite{Leghtas2013p023849} and references therein for approaches under the RWA.
Thus, we discuss how these trapped states allow for the creation and storage of highly entangled states between two atoms in the presence of an environment.

\section{Wootters concurrence in eigenstates of the two-qubit Dicke model} \label{sec:S2}

We are interested in two particular realizations of the two-qubit Rabi model; one is the Dicke model:
\begin{eqnarray}
\hat{H}_{1} = \omega \hat{n} + \omega_{0} \hat{S}_{z} + g \left( \hat{a}^{\dagger} + \hat{a} \right) \hat{S}_{x}, \label{eq:Dicke}
\end{eqnarray}
where the two-qubit operators are defined as $\hat{S}_{j} = \sum_{k} \hat{s}_{j}^{(k)}$ and the operators for the $k$th qubit are half the Pauli matrices, $\hat{s}_{j}^{(k)} = \hat{\sigma}_{j}^{(k)} / 2$, related to the ensemble of identical qubits of transition frequency $\omega_{0}$.
The boson field of frequency $\omega$ is described through the creation (annihilation), $\hat{a}^{\dagger}$ ($\hat{a}$), and number, $\hat{n} = \hat{a}^{\dagger} \hat{a}$, operators.
The Dicke model in (\ref{eq:Dicke}) describes the basic configuration of identical Rydberg atoms inside a high-Q cavity, solid-state qubits coupled to a strip line resonator or trapped ions in the same vibrational mode \cite{Garraway2011p1137}.
It is straightforward to realize that the corresponding Hilbert space can be split into triplet, $\left\{ \vert \Phi_{\pm, n} \rangle, \vert \Psi_{+,n} \rangle  \right\}$, and singlet, $\left\{ \vert \Psi_{-,n} \rangle \right\}$, sectors where
\begin{eqnarray}
\vert \Phi_{\pm, n} \rangle = \frac{1}{\sqrt{2}} \vert n \rangle \left( \vert g,g \rangle \pm \vert e,e \rangle \right),\\
\vert \Psi_{\pm,n} \rangle = \frac{1}{\sqrt{2}} \vert n \rangle \left( \vert g,e \rangle \pm \vert e,g \rangle \right).
\end{eqnarray}
We have discussed in the past a more general model of (\ref{eq:Dicke}), where the qubits are non-identical \cite{Chilingaryan2013p335301}, and will use our previously obtained knowledge in the present case. 
The Dicke model conserves parity and the sectors of the Hilbert space can be further subdivided into two parity subspaces each. 
In the singlet sector a particular eigenstate where $n$ is odd belongs to the positive parity subspace and those with even photon number $n$ belong to the negative parity subspace.
These singlet states are eigenstates of the Dicke model in (\ref{eq:Dicke}) with the $\lambda_{n}= \omega n$ as eigenvalues; their concurrence is maximal as they belong to the Bell basis, and the action of all orbital angular momentum operators over them is null, $\hat{S}_{j} \vert \Psi_{-,n} \rangle = 0$ with $j=x,y,z$.
Thus, even with the addition of driving, $\Omega_{y} \hat{S}_{y}$ or $\Omega_{x} \hat{S}_{x}$, they continue to be eigenstates.
Furthermore, the inclusion of dipole-dipole interactions, $\delta_{j} \hat{s}^{(1)}_{j}\hat{s}^{(2)}_{j}$ with $j=x,y,z$, only modifies the eigenvalue, such that $\lambda_{n}= \omega n - \sum_{j} \delta_{j} /4$, due to the fact that $\hat{s}^{(1)}_{j}\hat{s}^{(2)}_{j} \vert \Psi_{-,n} \rangle = - \vert \Psi_{-,n} \rangle$.
Thus, these eigenstates will become trapping states in the presence of an environment and driving, as we will show below.

Now, we can conjecture the prediction of three other classes of eigenstates, belonging to the triplet sector, via the Peres suggestion for finding new conserved quantities \cite{Peres1984p1711}.
In other words, a symmetry in any given system should manifest in the spectrum; i.e. we should be able to label the energies with quantum numbers. 
In the case of the JC model the symmetries are parity and number of excitations, but in the Dicke model only parity commutes with the Hamiltonian and the other symmetry is unknown.
We try to find an extra symmetry for the two-qubit Dicke model by exploring entanglement measures of its eigenstates in the triplet sector.
Note that any eigenvector in this sector can be written as
\begin{eqnarray}
\vert \psi_{+,\lambda} \rangle &=& \sum_{j} \left( c^{(+,\lambda)}_{0,j} \vert 2j,ee \rangle +  c^{(+,\lambda)}_{1,j} \vert \Psi_{+,2j+1} \rangle + c^{(+,\lambda)}_{2,j} \vert 2j,gg \rangle  \right), \label{eq:Ev1} \\
\vert \psi_{-,\lambda} \rangle &=& \sum_{j} \left( c^{(-,\lambda)}_{0,j} \vert 2j+1,ee \rangle +  c^{(-,\lambda)}_{1,j} \vert \Psi_{+,2j} \rangle + c^{(-,\lambda)}_{2,j} \vert 2j+1,gg \rangle  \right), \label{eq:Ev2} 
\end{eqnarray}
for positive and negative parity subspaces, in that order.
It is straightforward to see that $\langle \hat{S}_{x} \rangle = \langle \hat{S}_{y} \rangle  = 0$, but the action of the operators $\hat{S}_{j} \vert \psi \rangle$ will not necessarily be zero.
Thus, these states will not survive the inclusion of driving or coupling the system to an environment.
Furthermore, the reduced two-qubit states are $X$-states of the form
\begin{eqnarray}
\hat{\rho} = \left( \begin{array}{cccc}
r_{ee} & 0 & 0 & r_{eg} \\
0 & r_{\Psi} & r_{\Psi} & 0 \\
0 & r_{\Psi} & r_{\Psi} & 0 \\
r_{eg}^{\ast} & 0 & 0 & r_{gg} \\
\end{array} \right), \label{eq:Xstate} 
\end{eqnarray}
where $r_{ee} = \sum_{j} \vert c_{0,j}^{(\pm, \lambda)}\vert^{2}$, $r_{\Psi} = \frac{1}{2} \sum_{j} \vert c_{1,j}^{(\pm, \lambda)} \vert^{2} $, $r_{gg} = \sum_{j} \vert c_{2,j}^{(\pm, \lambda)} \vert^{2}$, and $r_{eg} = \sum_{j}  c_{0,j}^{(\pm, \lambda)} c_{2,j}^{(\pm, \lambda)\ast}$.
Here, it is straightforward to calculate the Wootters concurrence  \cite{Wootters1998p2245}, $C_{n} = \max(0, \epsilon_{1}- \epsilon_{2}-\epsilon_{3}-\epsilon_{4})$ where the real values $\epsilon_{j}$ are the square roots of the eigenvalues of $\tilde{\rho} \rho$  in decreasing order, $\epsilon_{1} > \epsilon_{2} > \epsilon_{3} > \epsilon_{4}$ with $\tilde{\rho} = \hat{\sigma}_{y}^{(1)} \hat{\sigma}_{y}^{(2)} \rho^{\ast} \hat{\sigma}_{y}^{(2)} \hat{\sigma}_{y}^{(1)}$; in our case one of the eigenvalues is zero and the other three are $4 \vert r_{\Psi} \vert^2$ and $r_{ee} r_{gg} + \vert r_{eg} \vert^{2} \pm 2 \vert r_{eg} \vert \sqrt{r_{ee} r_{gg}}$.

\begin{figure}[ht]
\centerline{\includegraphics[scale=1]{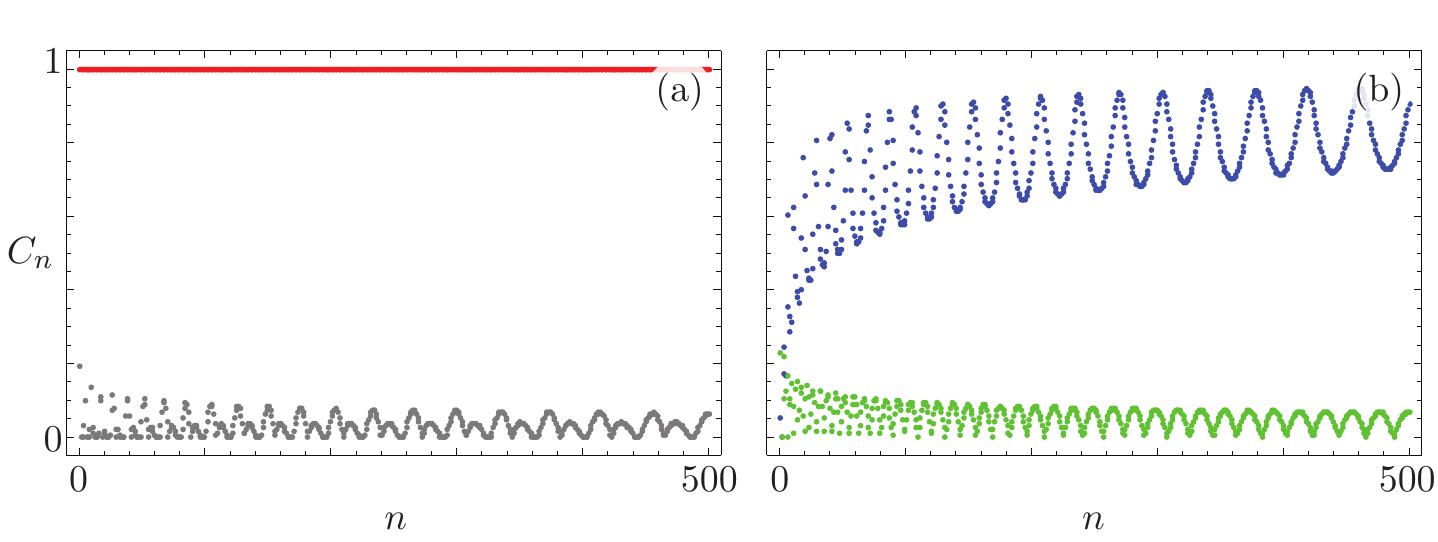}}
\caption{Wootters concurrence versus eigenvalues: (a) $\lambda_{4n+3}$ (light gray) and $\lambda_{4n+2}$ (dark red); (b) $\lambda_{4n+1}$ (light green) and $\lambda_{4n}$ (dark blue). The first $2000$ eigenstates are explored for the  positive parity subspace of the two-qubit Dicke model in (\ref{eq:Dicke}) on resonance in the ultra-strong coupling regime; i.e., $\omega_{0} = \omega$ and $g = 1.1 \omega$ with $\omega = 1$. The proper system was calculated using matrices of dimension $4000$ which provides a convergence error of $\delta \lambda_{n}\in [10^{-15},10^{-11}]$ and $\Delta V_{n} \in [10^{-17},10^{-15}]$. The red dots with $C_{n}=1$ in (a) correspond to the trapping states in the singlet sector.} \label{fig:Fig1}
\end{figure}

Figure \ref{fig:Fig1} shows the value of the concurrence for the first $2000$ eigenstates of the Dicke model in (\ref{eq:Dicke}) with parameters $\omega_{0} = \omega$, $g = 1.1 \omega$, and $\omega=1$; i.e., the qubit and the field frequencies are on resonance and the coupling corresponds to the USC regime.
The numerical diagonalization was performed in the whole positive parity subspace spanned by $\{ \vert 2k, e, e \rangle, \vert 2k, g, g \rangle , \vert 2k + 1, g, e \rangle, \vert 2k+1, e, g \rangle \}$ with $k=0,1,2,\ldots$ following \cite{Chilingaryan2013p335301}.
The subspace size, $S$, was increased until a convergence error, $\delta \lambda_{n}(S) =  \vert \lambda_{n}(S) - \lambda_{n}(S+1) \vert$, where $\lambda_{n}(S)$ is the $n$th eigenvalue for a matrix of size $S$ and $\Delta V_{n}(S) = 1- \vert \langle V_{n}(S) \vert V_{n}(S+1) \rangle \vert$, where $\vert V_{n}(S) \rangle$ is the $n$th eigenstate for a matrix of size $S$, of less than $10^{-10}$ was obtained; in the case shown, the size was $S=4000$ and the convergence errors were $\delta \lambda_{n} \in [10^{-15},10^{-11}]$ and $\Delta V_{n} \in [10^{-17},10^{-15}]$.
We explored an evenly distributed sample of 5000 coupling parameters in the range  $g \in \left[0.05,5\right] \omega$ on resonance, $\omega_{0}= \omega$, and another sample of 5000 with random coupling and detuning,  $g \in \left[0.05,5\right]\omega$ and $\omega_{0} \in \left[0.5, 1.5\right] \omega$.
Something equivalent can be done in the negative parity subspace and one obtains similar results. 
The common occurrence was finding four discernible clusters in the concurrence for the qubit part of the eigenstates;  the four clusters can be seen in $\hat{S}_{z}$ but are not so well defined, as they overlap. 
In the weak coupling regime, $g \sim 0.05$, the concurrences for triplet eigenstates overlap but the different clusters can be seen in the mean value of the energy difference, $\hat{S}_{z}$, and in the mutual quantum information of the qubits, to be described in the following section.
The trends in the concurrence point to a division of the spectrum of the model into four sections for each parity subspace: one regular section given by the singlet sector with constant spacing between continuous elements equal to $s_{1}=2 \omega$ and three belonging to the triplet sector where the distances between the eigenvalues are fairly regular, $s_{1} \sim 2 \omega$.
In the case presented in figure \ref{fig:Fig1} the regular sector was $\left\{ \lambda_{4n+2}\right\}$ with $n\ge0$ and the distance is given by $s_{1}=\lambda_{4n+6} - \lambda_{4n+2} = 2$, while the fairly regular sectors are  $\left\{ \lambda_{4n+1}\right\}$, $\left\{\lambda_{4n+3}\right\}$ and $\left\{\lambda_{4n+4} \right\}$ where the eigenvalue distances are $s_{1}=\lambda_{4n+j+4} - \lambda_{4n+j} \sim 2$ with $j=0,1,3$.
In addition, we calculated the mean average for the average nearest-neighbor spacing for both the numerical samples for the on-resonance and the off-resonance two-qubit Dicke model. 
Here we did not distinguish the singlet sector beforehand and just calculated the average fourth nearest-neighbor spacing, which happens to be equivalent to finding the average nearest neighbor by selecting the spectral classes.
The numerical results point to a fairly regular spectrum with spacings given by $s_{1} \sim 2 \omega$ as shown in Table \ref{tab:Tab1}.
Thus, thanks to the Peres criterion we can conjecture that the spectrum for the two-qubit Dicke model is fairly regular, as there are always three eigenvalues in the range $(\omega n, \omega (n+2))$ where $n$ is even or odd depending on the parity subspace.
In other words, there may exist an extra symmetry or partial symmetry \cite{Leviathan2011p93} beyond parity in the model.

\begin{table}
\centerline{ \begin{tabular}{c | c | c}
$\langle s_{1} \rangle$  & on-resonace & off-resonace \\
\hline
$\left\{ \lambda_{4n+1} \right\}$ & $2.00723 \pm 7.22187 \times 10^{-3}$ & $2.00719 \pm 7.16537 \times 10^{-3}$ \\
$\left\{ \lambda_{4n+2} \right\}$ & $2.00646 \pm 6.99162 \times 10^{-3}$ & $2.00641 \pm 6.93838 \times 10^{-3}$ \\
$\left\{ \lambda_{4n+3} \right\}$ & $2.00582 \pm 6.66601 \times 10^{-3}$ & $2.005476 \pm 6.62855 \times 10^{-3} $\\
$\left\{ \lambda_{4n+4} \right\}$ & $ 2.00532 \pm 6.26484 \times 10^{-3}$& $2.00524 \pm 6.24573 \times 10^{-3}$\\
\end{tabular}}
\caption{Mean average of the average spacing between continuous elements for the four subsets of the positive parity subspace spectrum of the two-qubit on- and off-resonance Dicke model. In each case, 2000 proper values for each one of the 5000 parameter  samples were considered. }  \label{tab:Tab1}
\end{table}

We tried to no avail to find such a symmetry.
We managed to arrive at an analytic expression for the eigenstate for the two-qubit Dicke model in the triplet sector,
\begin{eqnarray}
\vert f_{2}^{(\pm, \lambda)} \rangle &=& \frac{ \omega \hat{n} + \omega_{0} - \lambda}{\omega \hat{n} - \omega_{0} - \lambda} ~\vert f_{0}^{(\pm, \lambda)} \rangle, \\
\vert f_{0}^{(\pm, \lambda)} \rangle &=& - \frac{g}{\sqrt{2}} \left[\frac{\omega \hat{n} - \omega_{0} - \lambda}{(\omega \hat{n} - \lambda)^2 -\omega_{0}^2} \right]
\left( \hat{a} + \hat{a}^{\dagger}    \right) \vert f_{1}^{(\pm, \lambda)} \rangle,
\end{eqnarray}
where it is only necessary to determine
\begin{eqnarray}
\left\{ \lambda - \omega \hat{n} + g^{2} \left( \hat{a} + \hat{a}^{\dagger} \right) \left[ \frac{ \omega \hat{n}- \lambda}{(\omega \hat{n} - \lambda)^2 -\omega_{0}^2} \right]  \left( \hat{a} + \hat{a}^{\dagger} \right) \right\} \vert f_{1}^{(\pm, \lambda)} \rangle = 0. 
\end{eqnarray}
We have rewritten (\ref{eq:Ev1}) and (\ref{eq:Ev2}) as
\begin{eqnarray}
\vert \psi_{\pm, \lambda} \rangle = \vert f_{0}^{(\pm, \lambda)}\rangle \vert ee \rangle + \vert f_{1}^{(\pm, \lambda)}\rangle \vert \Psi_{+} \rangle + \vert f_{2}^{(\pm, \lambda)}\rangle \vert gg \rangle.
\end{eqnarray} 
In other words, we only need to determine two sets of coefficients, $\left\{ c_{1,j+1}^{(\pm, \lambda)} \right\}$, that obey a three-term recurrence relation,
\begin{eqnarray}
\alpha_{j}^{(\pm)}(\lambda) c^{(\pm,\lambda)}_{1,j-1} + \beta_{j}^{(\pm)}(\lambda) c^{(\pm,\lambda)}_{1,j} + \alpha_{j+1}^{(\pm)}(\lambda) c^{(\pm,\lambda)}_{1,j+1} = 0
\end{eqnarray} in each parity subspace with
\begin{eqnarray}
\alpha_{j}^{(+)}(\lambda) &=& \frac{g^{2} \sqrt{2j (2j+1)}  \left( 2j \omega - \lambda \right)}{\left(  2 j \omega - \lambda \right)^2 - \omega_{0}^{2} }, \\ \alpha_{j}^{(-)}(\lambda) &=& \frac{g^{2} \sqrt{2j (2j-1)}  \left[(2j-1) \omega - \lambda\right]}{\left[ (2j -1) \omega - \lambda \right]^2 - \omega_{0}^{2} }, \\
\beta^{(+)}_{j}(\lambda) &=& \lambda - (2j+1) \omega + \sqrt{\frac{2j+1}{2j}} ~\alpha_{j}^{(+)}(\lambda) + \sqrt{\frac{2j+2}{2j+3}} ~\alpha_{j+1}^{(+)}(\lambda), \\
\beta^{(-)}_{j}(\lambda) &=& \lambda - 2j \omega + \sqrt{\frac{2j-1}{2j}} ~\alpha_{j}^{(-)}(\lambda) + \sqrt{\frac{2j+1}{2j+2}} ~\alpha_{j+1}^{(-)}(\lambda).
\end{eqnarray} 
Note that the coefficients vanish, $c^{(\pm,\lambda)}_{1,j} =0$, for large values of $j$ as compared to the eigenvalues and frequencies, $j \gg \lambda, \omega, \omega_{0}$.
All the analytic recurrence relations were confirmed by our numerical eigenstates. 
These results reduce and simplify those provided in \cite{Chilingaryan2013p335301} and are equivalent to those presented more recently in \cite{Peng2013}; e.g., the spectra of the system reduce to two degenerate forced oscillator chains for cases well into the deep strong coupling regime, $g \gg \omega_{0}$, where it is possible to fully describe the dynamics of the system \cite{Chilingaryan2013p335301}.
We performed a statistical analysis of the boson field distributions, e.g. examining the Mandel Q parameter which seems to be always positive, but we gained no insight from these results.

\section{Mutual quantum information for eigenstates in a variation of the Dicke model}

\begin{figure}[ht]
\centerline{\includegraphics[scale=1]{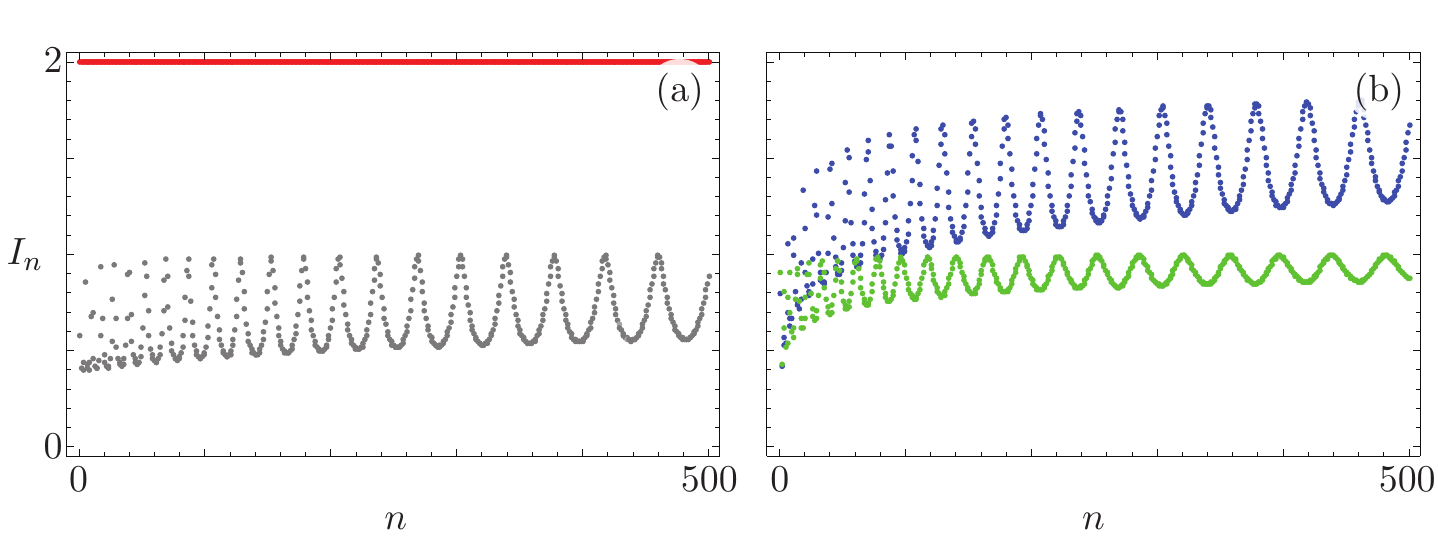}}
\caption {Mutual quantum information  versus eigenvalues: (a) $\lambda_{4n+3}$ (light gray) and $\lambda_{4n+2}$ (dark red); (b) $\lambda_{4n+1}$ (light green) and $\lambda_{4n}$ (dark blue). The first $2000$ eigenstates are explored for the  positive parity subspace of the two-qubit variation of the Dicke model in (\ref{eq:Model2}) on resonance in the ultra-strong coupling regime; i.e., $\omega_{0} = \omega$ and $g = 1.1 \omega$ with $\omega = 1$. The proper system was calculated using matrices of dimension $4000$ which provides a convergence error of $\delta \lambda_{n}\in [10^{-15},10^{-11}]$ and $\Delta V_{n} \in [10^{-17},10^{-15}]$. The dark red dots in (a) with $I_{n}=2$ correspond to the trapping states  $\vert \Phi_{-,n} \rangle$.}  \label{fig:Fig2}
\end{figure}

We are interested in showing that the mutual quantum information for two qubits is also a good test function for trends in the spectrum. 
For that reason, the second realization of the two-qubit Rabi model that we are interested in is a variation of the two-qubit Dicke model,
\begin{eqnarray}
\hat{H}_{2} = \omega \hat{n} + \omega_{0} \left( \hat{s}_{z}^{(1)} - \hat{s}_{z}^{(2)} \right) +  g \left( \hat{a}^{\dagger} + \hat{a} \right) \hat{S}_{x}. \label{eq:Model2}
\end{eqnarray}
Mathematically, this is just an unitary rotation over the second qubit around its $\sigma_{x}^{(2)}$ axis, $\hat{R} = e^{i \pi s_{x}^{(2)}}$.
Physically, this may be obtained from a two-trapped-ion scheme driven by a series of stationary lasers \cite{MoyaCessa2012p229}. 
Everything that we said before holds for this system, with the difference that the isolated eigenstate belongs to the triplet of the Bell basis, $\vert \Phi_{-,n} \rangle$.
They fulfill $\left( \hat{s}_{z}^{(1)} - \hat{s}_{z}^{(2)} \right) \vert  \Phi_{-,n} \rangle = 0$, $\hat{S}_{x} \vert  \Phi_{-,n} \rangle = 0$, $\hat{S}_{y} \vert  \Phi_{-,n} \rangle = -i \vert  \Psi_{+,n} \rangle $ and $\hat{S}_{z} \vert  \Phi_{-,n} \rangle = - \vert \Phi_{+,n}\rangle$.
So, they will be trapping states as long as the driving and dissipation do not involve $\hat{S}_{y}$ nor $\hat{S}_{z}$.
The quantum mutual information of the reduced qubit states, $I_{n} = \hat{S}^{(1)}_{vN} + \hat{S}^{(2)}_{vN} - \hat{S}_{vN}$ where the operators $\hat{S}_{vN}$ and  $\hat{S}^{(j)}_{vN}$ are the von Neumann entropy for the X-state and qubit $j=1,2$ reduced density matrices, can be easily calculated from the eigenstates of $X$-states as given in \cite{Girolami2011p052108}; 
the $\vert \Phi_{-,n} \rangle$  states  yield the maximum value of 2.
Again, we can intuit four classes of eigenstates; actually, the trends for the quantum mutual information are better defined than those for the concurrence (figure \ref{fig:Fig2}).
The numerical diagonalization was performed as described before.

\section{Trapping in the presence  of an environment} \label{sec:S3}

While the discussion about an adequate master equation describing the open dynamics of the Dicke model is still open \cite{Shirai2012}, let us discuss a typical environment scenario where the qubits couple to multiple boson field modes; this means that the extra terms
\begin{eqnarray}
\hat{H}_{RQ} = \sum_{j} \hat{b}_{j}^{\dagger} \hat{b}_{j}  + \sum_{j} \left( \hat{b}_{j}^{\dagger} + \hat{b}_{j} \right) \hat{S}_{x}
\end{eqnarray}
are added to (\ref{eq:Dicke}) and (\ref{eq:Model2}).
Here the multi-mode reservoir is described through the creation (annihilation) operators $\hat{b}_{j}^{\dagger}$ ($\hat{b}_{j}$) and these field modes are equivalent to having a non-ideal cavity or strip line resonator in cavity or circuit QED, or a real world driving laser with a linewidth  in ion-trap-QED.
Then, the trapping states have the form 
\begin{eqnarray}
\vert \Phi_{n} \rangle = \frac{1}{\sqrt{2}} \vert n \rangle_{a} \vert \left\{ m \right\} \rangle_{b}   \left( \vert g,e \rangle - \vert e,g \rangle \right), \label{eq:TSDickeDis}
\end{eqnarray}
for the two-qubit Dicke model, and
\begin{eqnarray}
\vert \Xi_{n} \rangle = \frac{1}{\sqrt{2}} \vert n \rangle_{a} \vert \left\{ m \right\} \rangle_{b}   \left( \vert e,e \rangle - \vert g,g \rangle \right), \label{eq:TSDickeDis}
\end{eqnarray}
for the variation of the two-qubit Dicke model. 
In both cases the set in the bosonic reservoir is defined as $\left\{m\right\}= \left\{m_{1}, m_{2}, \ldots \right\}$.
In other words, these states will not evolve.
If we also consider another typical source of noise coming from a non-ideal leaky cavity or strip line resonator for cavity and circuit QED setups or thermal radiation from the environment in an ion trap, we need to add the terms
\begin{eqnarray}
\hat{H}_{RC} = \sum_{j} \hat{c}_{j}^{\dagger} \hat{c}_{j}  + \sum_{j} \left( \hat{c}_{j}^{\dagger} + \hat{c}_{j} \right) \left( \hat{a}_{j}^{\dagger} + \hat{a}_{j} \right).
\end{eqnarray} 
Again, the qubit part of the state will not decay; it will be trapped.
Thus, one may intuit that in the presence of losses, a slightly excited qubit state will decohere and be trapped into these states while the field continues to decohere towards the coherent vacuum.

Note that it is possible to find a class of trapping states for the Dicke model in (\ref{eq:Dicke}) for qubit ensembles with even number of components, $N_{q} = 2k$, as
\begin{eqnarray}
\vert \phi_{n,k} \rangle =  \frac{1}{2^{k}} \vert n \rangle \prod_{(p,q)}  \left( \vert g,e \rangle_{(p,q)} - \vert e,g \rangle_{(p,q)} \right),
\end{eqnarray}
where the pairs $(p,q)$ cover all possible pairs of qubits without repeating a component; e.g., for four qubits we will have $\vert \Phi_{n}^{(1)} \rangle =  \frac{1}{2} \vert n \rangle  \left( \vert g,e \rangle_{(1,2)} - \vert e,g \rangle_{(1,2)} \right)\left( \vert g,e \rangle_{(3,4)} - \vert e,g \rangle_{(3,4)} \right)$, $\vert \Phi_{n}^{(2)} \rangle =  \frac{1}{2} \vert n \rangle  \left( \vert g,e \rangle_{(1,3)} - \vert e,g \rangle_{(1,3)} \right) \left( \vert g,e \rangle_{(2,4)} - \vert e,g \rangle_{(2,4)} \right)$, $\vert \Phi_{n}^{(3)} \rangle =  \frac{1}{2} \vert n \rangle  \left( \vert g,e \rangle_{(1,4)} - \vert e,g \rangle_{(1,4)} \right) \left( \vert g,e \rangle_{(2,3)} - \vert e,g \rangle_{(2,3)} \right)$ and all linear combinations of these.
These states will share the characteristic of  $\vert \Psi_{-,n} \rangle$ that  the action of all orbital angular momentum operators over them is null. 
Thus, the qubit ensemble part will not evolve even under driving and losses, as we showed for the two-qubit case.
Following an equivalent recipe, it is straightforward to construct trapping states for qubit ensembles of even size, $N_{q} = 2 k$, for the variation of the Dicke model in (\ref{eq:Model2}).
This corresponds to the Hamiltonian
\begin{eqnarray}
\hat{H}_{2} = \omega \hat{n} + \omega_{0} \sum_{(p,q)} \left( \hat{s}_{z}^{(p)} - \hat{s}_{z}^{(q)} \right) +  g \left( \hat{a}^{\dagger} + \hat{a} \right) \hat{S}_{x},
\end{eqnarray}
leading to trapping states of the form
\begin{eqnarray}
\vert \psi_{n,k} \rangle =  \frac{1}{2^{k}} \vert n \rangle \prod_{(p,q)}  \left( \vert e,e \rangle_{(p,q)} - \vert g,g \rangle_{(p,q)} \right).
\end{eqnarray}
Again, the pairs $(p,q)$ cover all possible pairs of qubits without repeating a component; e.g. if $\hat{H}_{2} = \omega \hat{n} + \omega_{0} \sum_{j=0}^{k-1} \left( \hat{s}_{z}^{(2j)} - \hat{s}_{z}^{(2j+1)} \right) +  g \left( \hat{a}^{\dagger} + \hat{a} \right) \hat{S}_{x}$ the trapping state will be given by $\vert \xi_{n,k} \rangle =  \frac{1}{2^{k}} \vert n \rangle \prod_{j}  \left( \vert e,e \rangle_{(2j,2j+1)} - \vert g,g \rangle_{(2j,2j+1)} \right)$ and all possible non-repeating iterations of index pairs of the form (even,odd) integers.
The difference here is that these states will only be impervious to the action of  $\hat{S}_{x}$.
Thus they will be trapping states as long as we do not add driving of the form $\Omega_{y} \hat{S}_{y}$ nor $\Omega_{z} \hat{S}_{z}$ .

\section{Conclusions} \label{sec:S6}

We have shown that the Wooters concurrence and quantum mutual information of the reduced qubit eigenstates point to the existence of four sets of normal modes of the two-qubit Dicke model, which may attest to the existence of a symmetry or partial symmetry in the model.
Numerical tests show that the corresponding four sets of eigenvalues are fairly regular, with eigenvalue spacings between elements close to two times the frequency of the field, i.e. a structure equivalent to that of an harmonic oscillator for each set. 
One of those eigenstate sets is a product state of a Fock state of the field and the singlet Bell state of the qubit ensemble with a regular spectrum with precise spacing of two times the field frequency. 
These eigenstates are maximal entangled states and are trapping states of the corresponding models coupled to an environment.
The trapping states given for the Dicke model are so robust that they continue trapped even in the presence of driving and dipole-dipole interactions.
Once the system reach these states, one can add several extra interactions that will not affect them; they will remain invariant, in principle, forever.
We explored 2000 proper values for each one of 5000 on-resonance and off-resonance model samples with parameter ranges $g \in \left[0.05,5\right]\omega$ and $\omega_{0} \in \left[0.5, 1.5\right] \omega$. 
We also explored a variation of the Dicke model where equivalent results were found regarding the spectrum, but the trapping states in this case were not so robust as they only resist a type of driving and dipole-dipole interaction.

\section*{Acknowledgment}
BMRL gratefully acknowledges valuable comments and suggestions from Francisco Soto-Eguibar.

\section*{References}


\providecommand{\newblock}{}

\end{document}